\documentclass[secnumarabic, amssymb, nobibnotes, aps]{revtex4-2}
\usepackage{epstopdf}
\usepackage{amsmath,amssymb,amsthm,amsfonts,mathrsfs,bm,verbatim}
\usepackage{graphicx,subfigure}
\usepackage{hyperref}
\usepackage{booktabs}
\newcommand{\bea}{\begin{eqnarray}}
\newcommand{\eea}{\end{eqnarray}}

\usepackage{float}

\def\bea{\begin{aligned}}
\def\ena{\end{aligned}}
\def\o{\omega}      
\def\d{\mathrm{d}}
\allowdisplaybreaks[4]

\begin{document}
	\title{Quasinormal modes and  stability  of higher dimensional rotating black holes under massive scalar perturbations }
	
	\author{Kai-Peng Lu\textsuperscript{1}}
	\author{Wenbin Li\textsuperscript{1}}
	\author{Jia-Hui Huang\textsuperscript{1,2,3}} \email{huangjh@m.scnu.edu.cn}
	
	\affiliation{$^{1}$ Key Laboratory of Atomic and Subatomic Structure and Quantum Control (Ministry of Education), School of Physics, South China Normal University, Guangzhou 510006, China}
	\affiliation{$^{2}$ Guangdong Provincial Key Laboratory of Quantum Engineering and Quantum Materials, South China Normal University, Guangzhou 510006, China}
	\affiliation{$^{3}$ Guangdong-Hong Kong Joint Laboratory of Quantum Matter, Frontier Research Institute for Physics, South China Normal University, Guangzhou 510006, China} 

	\begin{abstract}
	We consider the stability of six-dimensional singly rotating Myers-Perry black holes under massive scalar perturbations. Using Leaver's continued fraction method, we compute the quasinormal modes of the massive scalar fields. All modes found are damped under the quasinormal boundary conditions. It is also found that long-living modes called quasiresonances exist for large scalar masses as in the four-dimensional Kerr black hole case. Our numerical results provide a direct and complement evidence for the stability of six-dimensional MP black holes under massive scalar perturbation.	
	
	\end{abstract}

	\maketitle

	\section{Introduction}
	Black holes  are important objects which are predicted by general relativity and other extended gravitational theories. A crucial problem about the black hole solutions is their stability under some perturbations.  The issue of the stability of black holes was first considered by Regge and Wheeler over 60 years ago \cite{Regge:1957td}. They showed that if a Schwarzschild black hole is slightly perturbed in the metric, the perturbation field will oscillate and decay over time. Thus, the Schwarzschild black hole is stable.  

Rotating black holes, however, may develop instabilities under certain conditions. In early 1970's, it was pointed that rotating energy might be extracted from a classical rotating black hole by observers outside the black hole  \cite{Penrose:1971uk,Bardeen:1972fi,Press:1972zz,Bekenstein:1973mi}. When an ingoing bosonic wave is scattered off a Kerr black hole, it is found that the amplitude of the outgoing wave is greater than the ingoing one, if the frequency of the wave $ \o $ satisfying the superradiant condition $  0 < \o < m \Omega_H $, where $ m $ is the azimuthal quantum number of the incident wave, and $ \Omega_H $ is the angular velocity of the black hole horizon \cite{Brito:2015oca}. When there is a mirror-like mechanism that makes the amplified wave be scattered back and forth between the mirror and the black hole, the background black hole geometry will become superradiantly unstable. This is the black hole bomb mechanism \cite{Press:1972zz,Brito:2015oca,Cardoso:2004nk,Herdeiro:2013pia,Degollado:2013bha}.

Asymptotically flat Kerr black hole is stable under massless scalar, electromagnetic or gravitational perturbations\cite{Press:1973zz,Teukolsky:1974yv}. However, it has been shown that a massive scalar field may cause superradiant instability since  the mass term of the perturbing field acts as a natural mirror\cite{Damour:1976jd, Detweiler:1980uk,Furuhashi:2004jk,Dolan:2007mj}. The superradiant (in)stability of asymptotically flat Kerr black holes under massive scalar and vector perturbation has been studied extensively in the literature \cite{Strafuss:2004qc,Konoplya:2006br,Cardoso:2011xi,Dolan:2012yt,Hod:2012zza,Hod:2014pza,
Aliev:2014aba,Hod:2016iri,Degollado:2018ypf,Xu:2020fgq,Huang:2019xbu,Lin:2021ssw,East:2017ovw,East:2017mrj}.

The (in)stability of higher-dimensional rotating black holes has also been studied in literature \cite{Konoplya:2011qq,Jung:2005cn,Jung:2005nf}. The higher-dimensional extensions of four-dimensional Kerr black holes are Myers-Perry (MP) black holes\cite{Emparan:2008eg, Myers:1986un} and  the extensions with a nonvanishing cosmological constant was found by Gibbons, L\"{u}, Page, and Pope \cite{Gibbons:2004js}. It is found that MP-AdS black holes generally suffer from superradiant instability.
A MP-AdS black hole with equal angular momenta in an odd number of dimensions (greater than five) is shown to be superradiantly unstable under gravitational perturbations \cite{Kunduri:2006qa}. In $D>4$ spacetime dimensions, a simply rotating MP-AdS black hole is shown to be superradiantly unstable against tensor-type gravitational perturbations when its angular momentum is larger than a critical value\cite{Kodama:2009rq}. Superradiant instability has also been proven for small singly rotating MP-AdS black holes in arbitrary dimensions under massive scalar perturbations \cite{Delice:2015zga}. Recently, the quasinormal modes of higher-dimensional singly rotating MP-dS black holes with non-minimally coupled scalar fields were analytically studied, and a near-extremal formula for quasinormal frequencies was derived \cite{Gwak:2019ttv}. Furthermore, it is found that MP-dS black hole with a single rotation is stable under a massive scalar perturbation\cite{Ponglertsakul:2020ufm}. 

Asymptotically flat MP black holes with equal angular momenta are found to be stable in five and seven spacetime dimensions and unstable in nine spacetime dimension against gravitational perturbation \cite{Dias:2010eu}. In $D>6$ spacetime dimensions, singly rotating MP black holes are found to be stable against tensor-type gravitational perturbations since an extensive search of quasinormal modes for these black holes did not indicate any presence of growing modes \cite{Kodama:2009bf,Berti:2003yr}. In \cite{Ida:2002zk}, the quasinormal modes of a massless scalar field were studied in five-dimensional MP black hole using the continued fraction
method. The numerical result shows that the singly rotating MP black hole is stable. Similarly, a six-dimensional singly rotating MP black holes are shown to be stable  under massless scalar perturbations \cite{Cardoso:2004cj}, even in the ultra-spinning regime \cite{Morisawa:2004fs}. In \cite{ Cardoso:2005vk}, the authors considered the stability of singly rotating MP black holes against massive scalar perturbations, qualitatively analyzed the effective potential in the radial equation of motion of the massive scalar field with the assumption that there are no stable orbits for $D>4$ MP black holes, and concluded that there is no potential well for supporting bound states and MP black holes are stable.   
In this paper, we will numerically study quasinormal modes of the massive scalar field in the six-dimensional singly rotating MP black hole with the continued fraction
method, and provide a direct and complement evidence for  the stability of six-dimensional singly rotating MP black holes.
	
	This paper is organized as follows. In Sec.~\ref{sec. 2}, we present the general equation of motion of a massive scalar field in a singly rotating MP black hole, and divide it into the angular equation and the radial equation. Furthermore, we study the asymptotic behaviors of the radial wave function under the quasinormal mode boundary conditions. In Sec.~\ref{sec. 3}, we rewrite the radial equation into a Schr\"{o}dinger-like equation and study the asymptotic behavior of the effective potential. In Sec.~\ref{sec. 4}, we describe the continued fraction method used here to compute the quasinormal modes of the massive scalar field. In Sec.~\ref{sec. 5}, we present our numerical results which show that there is no unstable mode. The final section is devoted to the conclusion.

	\section{The singly rotating black hole and a massive scalar perturbation} \label{sec. 2}
	
	For four-dimensional Kerr black hole, there is only one angular momentum parameter. However, for D-dimensional MP black holes($D>4$),  there are  $\lfloor\frac{D-1}{2}\rfloor$  angular momentum parameters. Here, we consider the singly rotating MP black hole with only one nonzero angular momentum parameter denoted by $a$.
	The metric of such a MP black hole is given in Boyer-Lindquist-type coordinates by  \cite{Myers:1986un,Cardoso:2004cj}
	\begin{equation} \label{2}
		\bea
		\d s^2 =& -\dfrac{\Delta-a^2 \sin^2\theta}{\Sigma} \d t^2-\dfrac{2a(r^2+a^2-\Delta)\sin^2\theta}{\Sigma}\d t \d \varphi+\dfrac{(r^2+a^2)^2-\Delta a^2 \sin^2\theta}{\Sigma} \sin^2\theta \d \varphi^2 \\
		&+\dfrac{\Sigma}{\Delta} \d r^2+\Sigma \d \theta^2+r^2 \cos^2\theta \d \Omega_n^2,
		\ena
	\end{equation}
	where
	\begin{align} \label{3}
		\Sigma &= r^2 + a^2 \cos^2 \theta,\\
		\Delta &= r^2 + a^2 - 2 M r^{1-n},
	\end{align}
	and $ \d \Omega_n^2 $ denotes the standard metric of the unit $ n $-sphere ($n=D-4$). The  mass and angular momentum of the black hole are respectively proportional to $ M $ and  $M a$ with some dimension-dependent coefficients. Without loss of generality, we assume $ M, a > 0 $.  When $D\geq 6$, equation $ \Delta = 0 $ has exactly one positive root for arbitrary $ a>0 $. There is no bound on $ a $, which means there is no extremal MP black hole in these cases. 

The equation of motion of a massive scalar perturbation field $ \Psi $ with mass $\mu$ in the MP black hole background is governed by following covariant Klein-Gordon equation	  
	\begin{equation}  \label{4}
		\nabla^\nu \nabla_\nu \Psi =\dfrac{1}{\sqrt{-g}} \dfrac{\partial}{\partial x^\mu} \left( \sqrt{-g} g^{\mu\nu} \dfrac{\partial}{\partial x^\nu} \Psi \right) = \mu^2 \Psi.
	\end{equation}
The above equation was shown to be separable \cite{Ida:2002zk}, and the solution with definite angular frequency of it can be decomposed as
	\begin{equation}
		\Psi=e^{-i \omega t+im \varphi} R(r)S(\theta)Y(\Omega),
	\end{equation}
	where $Y(\Omega)$ are generalized scalar spherical harmonics on $ n $-sphere with eigenvalues $-j(j+n-1) \ (j=0,1,2,...)$. Substitute this solution into \eqref{4}, we can obtain the angular equation for the angular function $S(\theta)$
	\begin{equation} \label{12}
		\dfrac{1}{\sin\theta \cos^n\theta} \dfrac{\d}{\d \theta} \left(\sin\theta \cos^n\theta \dfrac{\d S}{\d \theta} \right)+ \left[ a^2 \left( \omega^2 - \mu^2 \right) \cos^2\theta -m^2 \csc^2\theta -j(j+n-1)\sec^2\theta +A_{kmj} \right]S=0,
	\end{equation}
	and radial equation for the radial function $R(r)$
	\begin{equation} \label{5}
	    r^{ -n } \dfrac{\d}{\d r} \left(r^n \Delta \dfrac{\d R}{\d r} \right)+U R=0,
	\end{equation}
	where
	\begin{align}
		U &=\dfrac{ \left[ \o \left( r^2 +a^2 \right) - m a \right]^2 }{ \Delta } - \left[ -2 a m \o + \o^2 a^2 + \mu^2 r^2 + \dfrac{j \left( j + n - 1 \right) a^2 }{ r^2 } + A_{kmj} \right], \\
		A_{kmj} &=\left( 2k + j + |m| \right) \left( 2k + j + |m |+ n + 1 \right). \quad (k=0,1,2,\cdots)
	\end{align}
	Here, the above separation constants  $ A_{kmj} $ is calculated in the limit $ a \o \rightarrow 0 $.  $k, j, m$ are all angular quantum numbers.

In this paper, we will consider the quasinormal modes of the massive scalar field. 	By imposing the quasinormal mode boundary conditions, which correspond to purely ingoing waves at the event horizon and purely outgoing waves at spatial infinity, we can obtain the following asymptotic behaviours for the radial function at the boundaries, 
	\begin{equation} \label{6}
		R \sim
		\begin{cases}
			\left( r - r_H \right)^{- i \sigma } & \quad r \rightarrow{ r_H } ,\\
			r^{ - ( n + 2 ) / 2 } e^{ i \sqrt{ \o^2 - \mu^2 } r } & \quad r \rightarrow{ \infty  },
		\end{cases}
	\end{equation}
	where
	\begin{equation}
		\sigma = \dfrac{ \left[ \left( r_H ^2 + a^2 \right) \o - m a \right] r_H }{ ( n - 1 ) \left( r_H ^2 + a^2 \right) + 2 r_H ^2 } 
	\end{equation}
	can be determined by the asymptotic solution of Eq.\eqref{5} near horizon.

	\section{The effective potential} \label{sec. 3}
	From now on, we will concentrate on the quasinormal modes of the massive scalar field in six-dimensional MP black hole, which may give us some idea about that in other higher dimensional MP black holes. When $D=6$, $ n=2 $.  The event horizon $ r_H $ of the six-dimensional MP black hole is given by
	\begin{equation}
		r_H=\dfrac{-3^{1/3} a^2+(9M+\sqrt{3} \sqrt{a^6+27 M^2})^{2/3}}{3^{2/3} (9M+\sqrt{3} \sqrt{a^6+27 M^2})^{1/3}}.
	\end{equation}

Defining the tortoise coordinate $ r_* $ by $ \dfrac{ \d r}{ \d r_* } = \dfrac{ \Delta }{ r^2 + a^2  } $ and a new radial function $ \psi (r) = \sqrt{ r^2 \left( r^2 + a^2 \right) } R(r) $, the radial equation \eqref{5} can be rewritten as
	\begin{equation}
		\dfrac{\d^2 \psi}{\d r^2}+(\omega^2-V)\psi=0,
	\end{equation}
	and the effective potential is 
	\begin{equation}
		V=\omega^2-\frac{ \Delta \left( r^2 + a^2 \right)^2 r U - \left( \Delta ^2 r \left(a^2-2 r^2\right)+\Delta  \left(a^2+r^2\right) \left(a^2+2 r^2\right) \Delta' \right) }{ r \left( a^2 + r^2 \right)^4 }.
	\end{equation}
Before we do the numerical calculation, we first consider the asymptotic behaviors of the effective potential and its derivative. The asymptotic behaviors of the effective potential $ V $ near the horizon and at spatial infinity are
	\begin{equation}
		\bea
		V (r \rightarrow{r_H})   & \rightarrow  \o^2 - \left( \o - \o_c \right)^2 , \\
		V (r\rightarrow{+\infty})  & \rightarrow  \mu^2 + \dfrac{2 + A_{kmj} +a^2 \left( \o^2 - \mu^2 \right)  }{ r^2 }+\mathcal{O} \left( \dfrac{1}{r^3} \right), 		
		\ena
	\end{equation}
	where
	\begin{equation}
		\o_c = \dfrac{ m a }{ r_H^2 + a^2 },
	\end{equation}
is the threshold of superradiant angular frequency.  The asymptotic behavior of the derivative of the effective potential $ V $ at spatial infinity is
	\begin{equation}
		V'(r \rightarrow {+\infty})  \rightarrow -\dfrac {2 \left(2 + A_{kmj} +a^2 \left( \o^2 - \mu^2 \right) \right)}{r^3}  + \mathcal{O} \left( \dfrac{1}{r^4} \right). 
	\end{equation}
The derivative of the effective potential is negative by considering the quasinormal mode boundary condition at spatial infinity.	Based on the asymptotic behavior of the effective potential, there is no trapping well at spatial infinity. This picture suggests that the six-dimensional MP black holes are stable. This analysis is similar to that in Ref.\cite{Cardoso:2005vk}.

	\section{Numerical Computation} \label{sec. 4}
	 In previous section, we analyze the behaviors of the effective potential and the analysis on the stability of the six-dimensional MP black holes against massive scalar perturbation. However, according to the asymptotic behaviors of the effective potential and its derivative at spatial infinity, we can only rigorously infer that there is no well near the spatial infinity. In fact, we don't know whether there is a potential well between the horizon and spatial infinity. So it is interesting to provide a direct numerical calculation to confirm that the quasinormal modes are stable.

To get the frequencies of the quasinormal modes $ \o_{Q} $, we take $M=1$, and choose a set of values of the scalar field mass parameter $\mu$, rotation parameter $ a $ and the angular quantum numbers $ k,\, j,\, m $, and then numerically solve the angular equation and radial equation simultaneously to obtain $A_{kjm}$ and $ \o_{Q} $. The imaginary parts of the quasinormal modes govern the growth or decay of the perturbation field. If $ \o_{Q} $ has a negative imaginary part, the modes will decay and the black hole is stable under the perturbation. Here, we use the continued fraction method to compute the quasinormal frequencies \cite{Leaver:1985ax,Berti:2009kk,Kokkotas:1999bd,Konoplya:2011qq}, which is described in the following two subsections.

	\subsection{The Angular equation}
We now consider the six-dimensional angular equation, which is Eq. \eqref{12} with $n=2$. The angular function can be assumed to have the following form \cite{Berti:2005gp},
	\begin{equation}
		S = \left( \sin \theta  \right)^{ |m| } \left( \cos \theta \right)^j \sum_{ k=0 }^{ \infty } a_k \left( \cos^2 \theta \right)^k .
	\end{equation}
	Substitute this form into Eq.\eqref{12} with $n=2$, one can obtain a three-term recursion relation,
	\begin{eqnarray} \label{13}
		&&\alpha_0^\theta a_1+ \beta_0^\theta a_0 =0\,, \nonumber \\
		&&\alpha_k^\theta a_{ k+1 }+ \beta_k^\theta a_k+ \gamma_k^\theta a_{ k-1 } =0 \, \quad ( k = 2, 3, \cdots),
	\end{eqnarray}
	where
	\begin{eqnarray}
		&& \alpha_k^\theta =  -2 \left( k + 1 \right) \left( 2j + 2k + 3  \right)\,, \nonumber \\
		&& \beta_k^\theta = \left( 2k + j + |m| \right) \left( 2k + j + |m |+ 3 \right) -A_{kjm} \,, \nonumber \\
		&& \gamma_k^\theta = a_*^2 \left( \mu_*^2 - \o_*^2 \right)\,.
	\end{eqnarray}
	Here, we use the dimensionless quantities $ \o_* := \o r_H,\, \mu_* = \mu r_H $ and $ a_* := a / r_H $. Now, the behavior of the system is simply determined by $ a_*,\, \mu_* $ and the angular indices $ k,\, j,\, m $. In addition, one can easily find that this recursion relation reduces to the massless case when $ \mu_* = 0 $. Furthermore, in the Schwarzschild limit $ ( a_* = 0 ) $, $ \gamma_k^\theta $ are zero for all $ k $, so the recursion will stop whenever $ A_{kjm} $ is such that $ \beta_k^\theta $ is zero for some $ k $, which gives $ A_{kjm} =\left( 2k + j + |m| \right) \left( 2k + j + |m |+ 3 \right) $.	
	For  given $ m,\, j,\, k,\, a_*,\, \mu_*,\, \text{and } \o_* $,  the separation constant $ A_{kjm} $ is a root of the continued fraction
	\begin{equation}
		0 = \beta_0^\theta - \frac{ \alpha_0^\theta \gamma_1^\theta}{ \beta_1^\theta - } \frac{ \alpha_1^\theta \gamma_2^\theta }{ \beta_2^\theta - } \frac{ \alpha_2^\theta \gamma_3^\theta }{ \beta_3^\theta - \cdots }.
	\end{equation}
	
	\subsection{The Radial equation}	
	When $ n > 1 $,	a solution of the radial equation  \eqref{6} can be expressed as:
	\begin{equation} \label{9}
		R = e^{ i \sqrt{ \o^2 - \mu^2 } r } \left( \frac{ r }{ r_H } \right) ^{ - (n+2) /2 + i \sigma }   \left( \frac{ r - r_H }{ r_H } \right) ^{ - i \sigma }  \sum_{ n = 0 }^{ \infty } b_n \left( \frac{ r - r_H }{ r } \right) ^n ,
	\end{equation}
	where $ b_0 $ is usually taken to be $ 1 $. When $ n=2 $, we rewrite the above radial function with the dimensionless quantities
	\begin{equation} \label{10}
		R = e^{ i \sqrt{ \o_*^2 - \mu_*^2 } x } x ^{ - 2 + i \sigma }   \left( x - 1 \right) ^{ - i \sigma }  \sum_{ n = 0 }^{ \infty } b_n \left( \frac{ x - 1 }{ x } \right) ^n .
	\end{equation}
	Here $ x = r / r_H $. Substitute the above expression into the radial equation \eqref{5}, one can get a nine-term recurrence relation,
	\begin{eqnarray} \label{11}
		&&\alpha_0 b_1+ \beta_0 b_0 =0\,, \nonumber \\
		&&\alpha_1 b_2+ \beta_1 b_1+ \gamma_1 b_0 =0\,, \nonumber \\
		&&\alpha_2 b_3+ \beta_2 b_2+ \gamma_2 b_1+ \delta_2 b_0 =0\,, \nonumber \\
		&&\alpha_3 b_4+ \beta_3 b_3+ \gamma_3 b_2+ \delta_3 b_1+ \epsilon_3 b_0 =0\,, \\
		&&\alpha_4 b_5+ \beta_4 b_4+ \gamma_4 b_3+ \delta_4 b_2+ \epsilon_4 b_1+ \zeta_4 b_0 =0\,, \nonumber \\
		&&\alpha_5 b_6+ \beta_5 b_5+ \gamma_5 b_4+ \delta_5 b_3+ \epsilon_5 b_2+ \zeta_5 b_1+ \theta_5 b_0 =0\, , \nonumber \\
		&&\alpha_6 b_7+ \beta_6 b_6+ \gamma_6 b_5+ \delta_6 b_4+ \epsilon_6 b_3+ \zeta_6 b_2+      \theta_6 b_1+ \rho_6 b_0 =0\, , \nonumber \\
		&&\alpha_n b_{ n+1 }+ \beta_n b_n+ \gamma_n b_{ n-1 }+ \delta_n b_{ n-2 }+ \epsilon_n b_{ n-3 }+ \zeta_n b_{ n-4 }+ \theta_n b_{ n-5 }+ \rho_n b_{ n-6 }+ \phi_n b_{ n-7 } =0 .\, \quad ( n = 7, 8, \cdots) \nonumber 
	\end{eqnarray}	
	We don't present the coefficients $\{\alpha_n,\beta_n,...,\phi_n\}$ here since they have rather complicated forms. By making Gaussian eliminations six times, we can reduce the nine-term recurrence relations \eqref{11} to a three-term recurrence relations, which is given by
	\begin{eqnarray} 
		&& \tilde\alpha_0 b_1 + \tilde\beta_0 b_0 = 0\,, \nonumber \\
		&& \tilde\alpha_n b_{ n+1 } + \tilde\beta_n b_n +\tilde\gamma_n b_{ n-1 } = 0 .\, \quad ( n = 2, 3, \cdots ) 
	\end{eqnarray}
	The details of the Gaussian elimination method can be found in \cite{Leaver:1990zz}. The series in \eqref{9} converges, and the $ r = \infty $ boundary condition is satisfied if, for a given set of the values of parameters $ \{ m, j, A_{kjm}, a_*, \mu_* \} $, the frequency $ \o_* $ is a root of the continued fraction equation
	\begin{equation}
		\tilde\beta_0-{\tilde\alpha_0\tilde\gamma_1\over\tilde\beta_1-}
		{\tilde\alpha_1\tilde\gamma_2\over\tilde\beta_2-}
		{\tilde\alpha_2\tilde\gamma_3\over\tilde\beta_3-} \cdots
		\equiv
		\tilde\beta_0-\frac{\tilde\alpha_0\tilde\gamma_1}{\tilde\beta_1-
			\frac{\tilde\alpha_1\tilde\gamma_2}
			{\tilde\beta_2-\frac{\tilde\alpha_2\tilde\gamma_3}
				{\tilde\beta_3-\cdots}}} =0 \,.
	\end{equation}

	\section{Numerical Results} \label{sec. 5}
\begin{figure}[tbh] 
		\centering
		\includegraphics[width=0.45\textwidth]{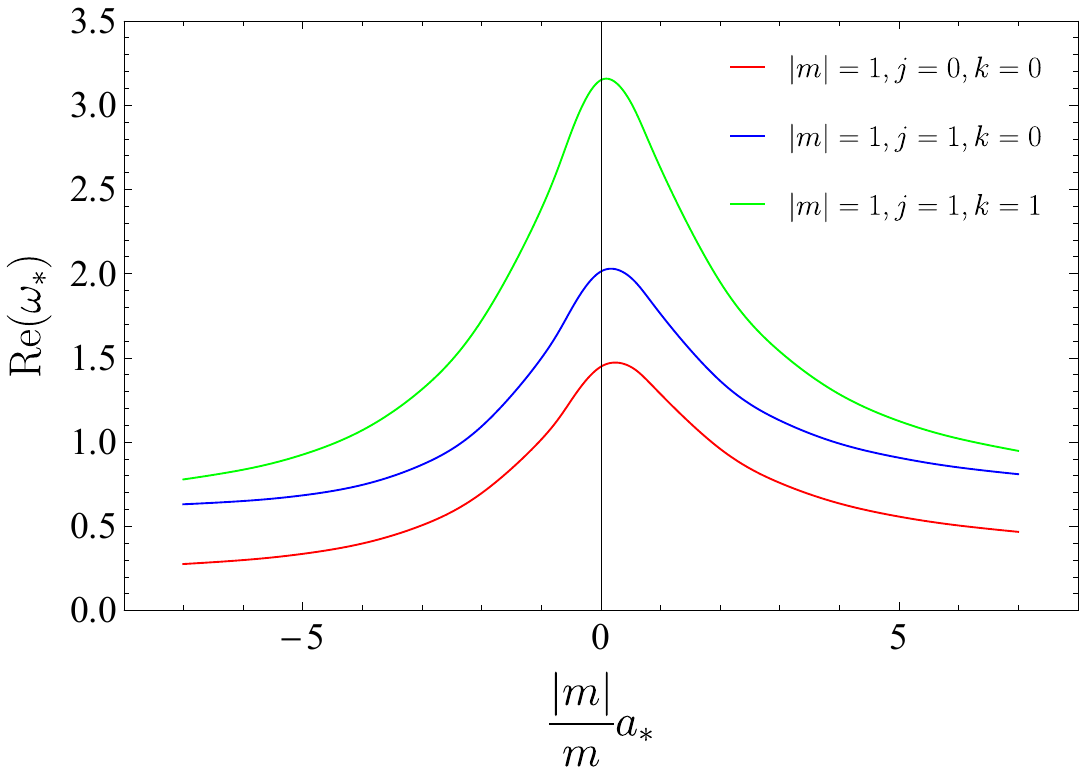}
		\includegraphics[width=0.45\textwidth]{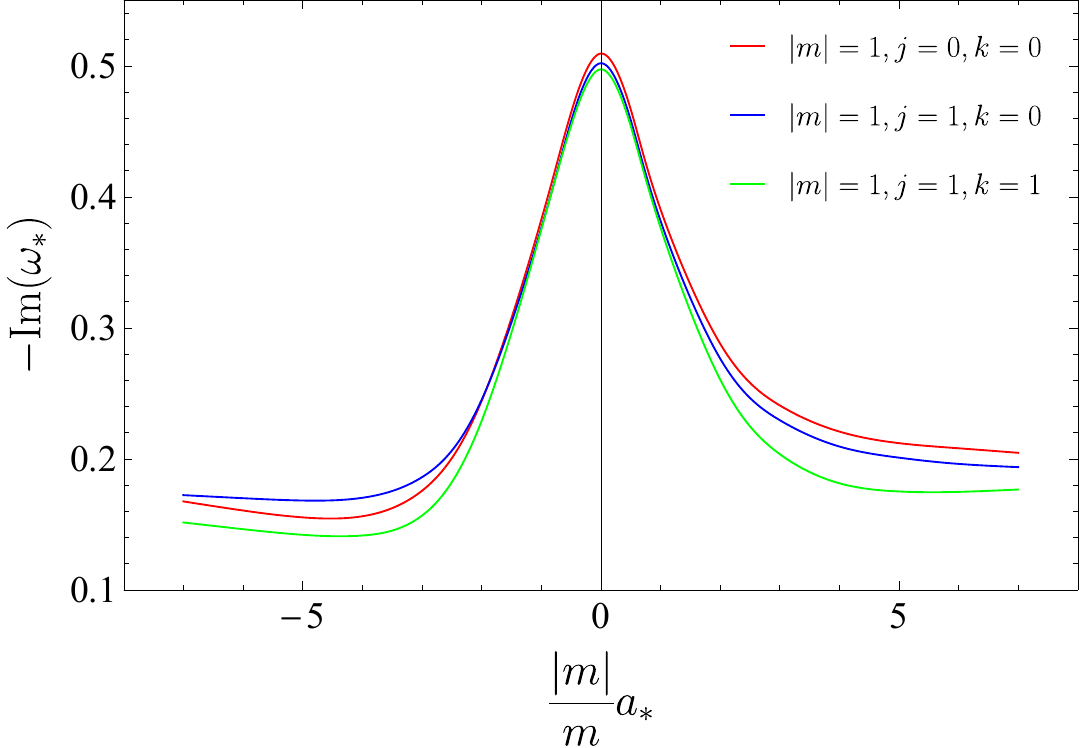}
		\caption{Fundamental quasinormal modes under the massless scalar perturbation.}
		\label{fig1} 
	\end{figure}

\begin{figure}[tbh] 
		\centering
		\includegraphics[width=0.40\textwidth]{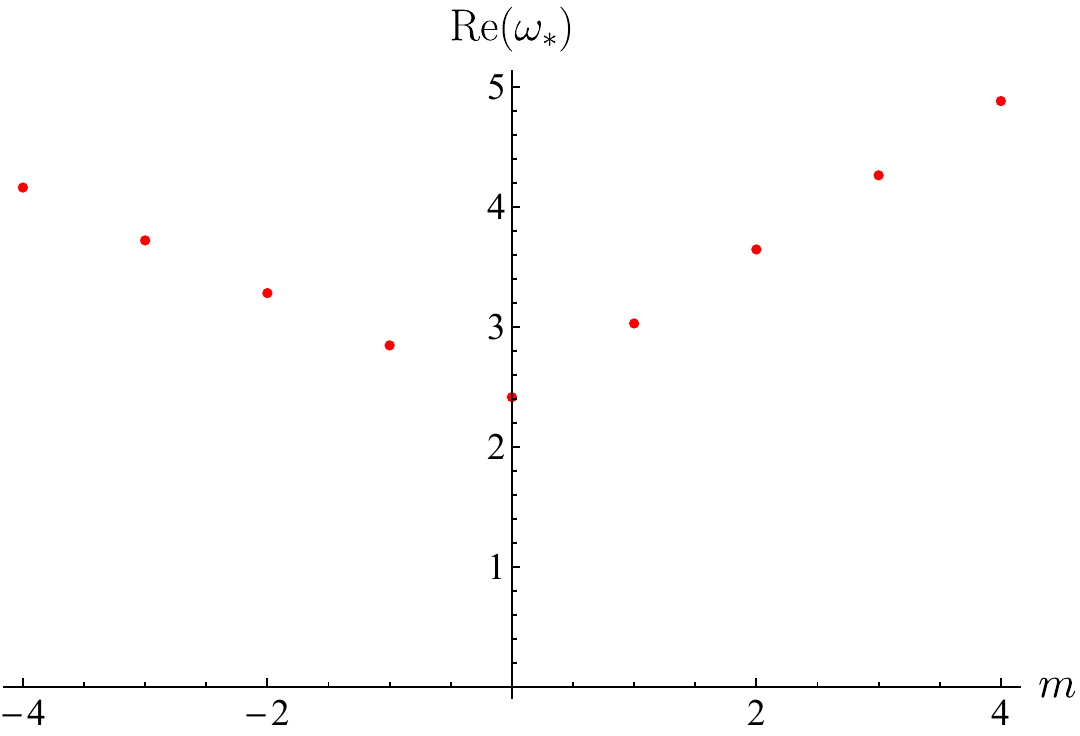}
        \hspace{5mm}
		\includegraphics[width=0.40\textwidth]{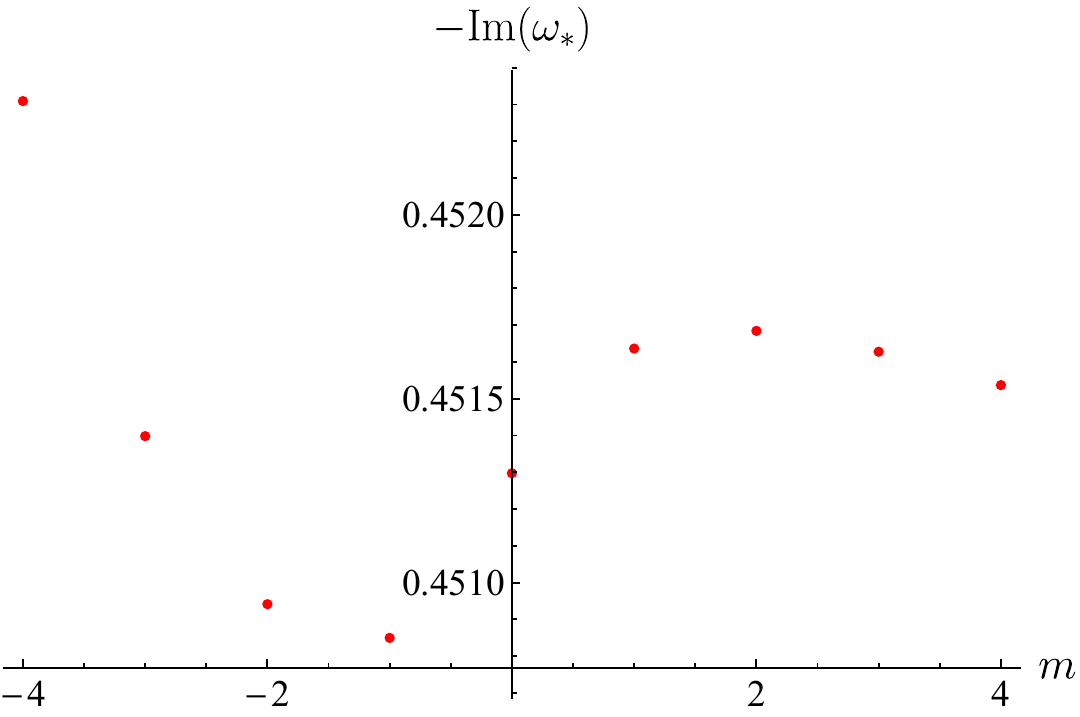}
		\caption{Re($ \o_* $) and Im($ \o_* $) as a function of azimuthal number $ m $ for $ a_* = 0.5,\, \mu = 0.3,\, j = 1,\, k = 1 $.}
		\label{fig2} 
	\end{figure}

\begin{figure}[tbh] 
		\centering
		\includegraphics[width=0.42\textwidth]{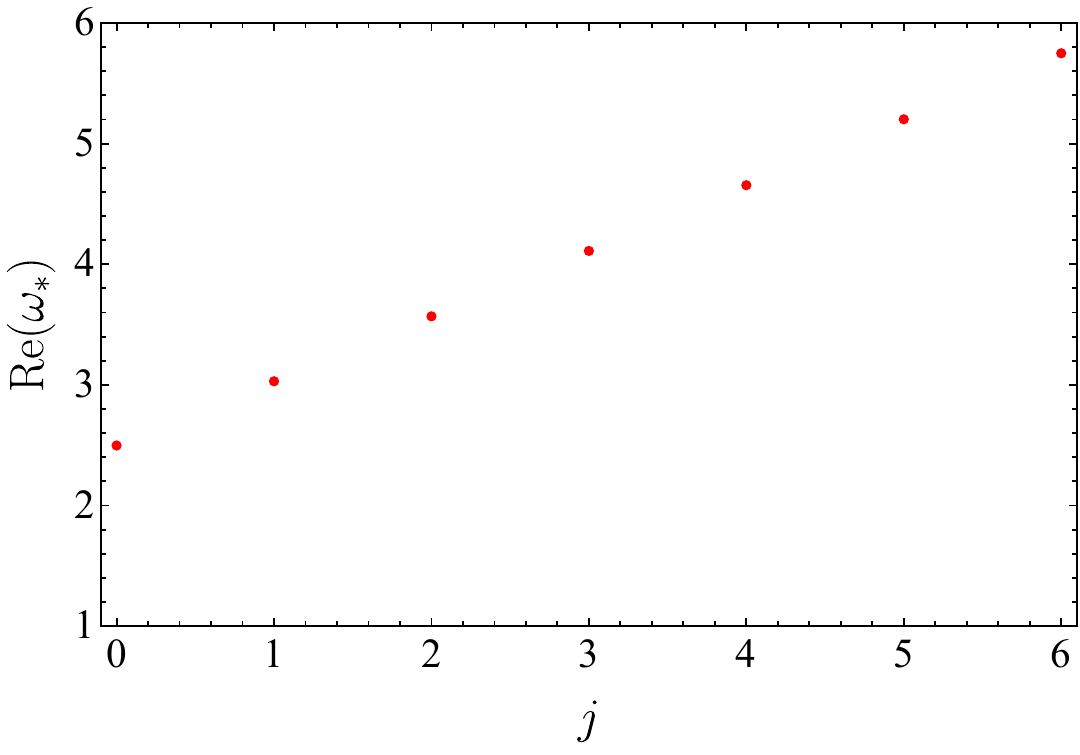}
        \hspace{5mm}
		\includegraphics[width=0.45\textwidth]{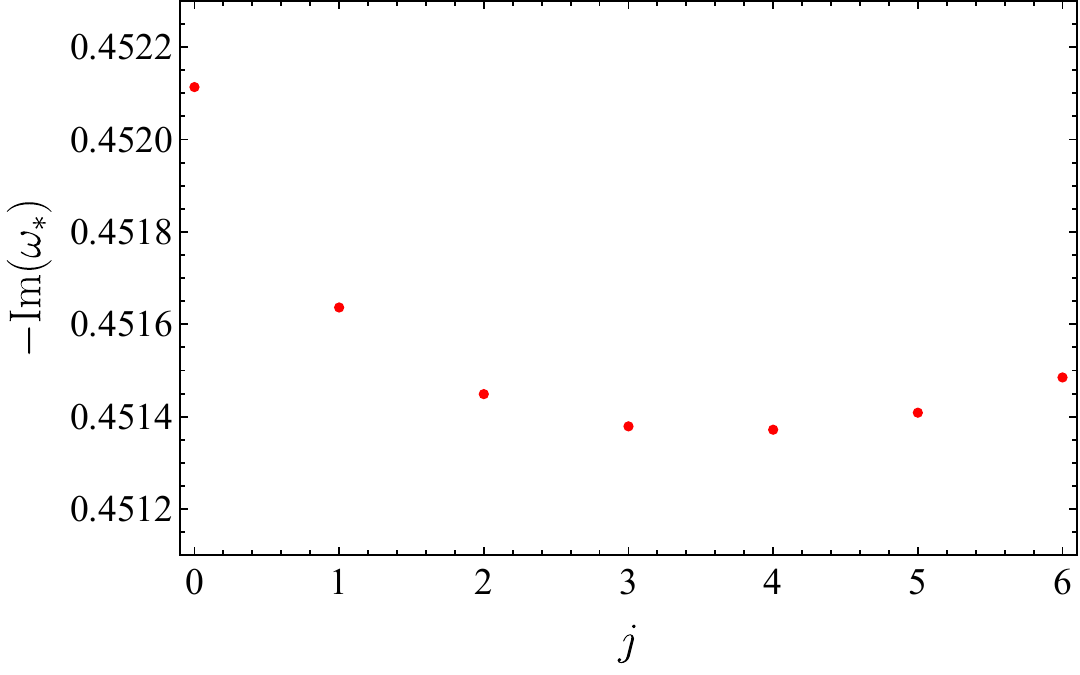}
		\caption{Re($ \o_* $) and Im($ \o_* $) as a function of $ j $ for $ a_* = 0.5,\, \mu = 0.3,\, m = 1,\, k = 1 $.}
		\label{fig3} 
	\end{figure}

\begin{figure}[tbh] 
		\centering
		\includegraphics[width=0.6\textwidth]{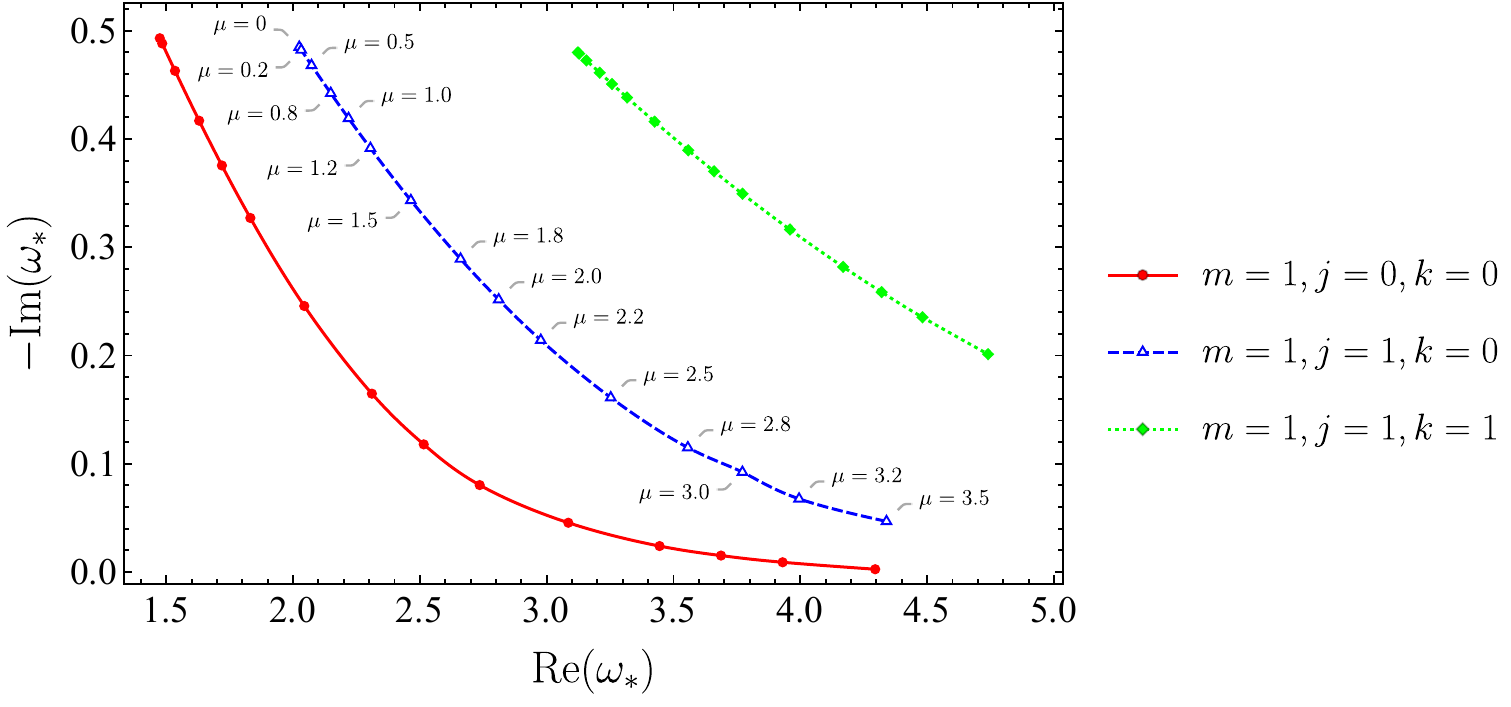}
		\caption{The real and imaginary parts of the fundamental quasinormal modes for different values of scalar field mass $ \mu $ and angular quantum numbers $ m, j, k$. In each line, there are fifteen points corresponding to different values of scalar masses. For simplicity, we just mark them in one line.  Here the dimensionless rotation parameter of the black hole is $ a_* = 0.3 $.}
		\label{fig4} 
	\end{figure}

	Using the method described above, we have computed numerous quasinormal frequencies in the six-dimensional MP black hole for different values of $ k, j, m, $ and $ \mu $. The numerical results are summarized in Fig.\eqref{fig1}, Fig.\eqref{fig2}, Fig.\eqref{fig3}, Fig.\eqref{fig4} and Table.\eqref{table.1}. To check our code, we have computed quasinormal frequencies of the massless scalar perturbation in Fig.\eqref{fig1}, and compared our results with the ones obtained previously \cite{Cardoso:2004cj}.  We find agreement between our results and the ones in \cite{Cardoso:2004cj}. 
	
	In Table.\eqref{table.1}, we calculate the quasinormal frequencies of the massive scalar perturbations with different masses  $ \mu=0.1, 0.2, 0.3 $. For the quasinormal frequencies and the rotation parameters, we present them using the corresponding dimensionless quantities defined in previous section.  We stop our computation when the imaginary part of $ \o_* $ changes slowly. 
 
 Table.\eqref{table.1} provides a direct evidence that the quasinormal modes of massive scalar perturbations are decaying modes. Furthermore, these results indicate that the damping rate, which is determined by  $ -\text{Im}(\o_*) $, slightly decreases as the mass $ \mu $ of the scalar perturbation increases, while the real part of $ \o_* $ slightly increases as $ \mu $ grows. This property is qualitatively the same as that of the four-dimensional Kerr black hole under a massive scalar perturbation \cite{Dolan:2007mj, Konoplya:2006br}. Quantitatively, it is noticeable that the imaginary part of $ \o_* $ in the six-dimensional MP black hole is significantly larger than that in the four-dimensional case \cite{Konoplya:2006br}, which implies faster damping and shorter lifetime in a higher dimension. This property is similar to that in the Schwarzschild black hole case \cite{Zhidenko:2006rs}.
	
	\begin{table}[tbh] 
		\centering
		\caption{Values of the fundamental quasinormal modes with $m = j = k = 1$ for different values of perturbation mass $ \mu $, and rotation parameter $ a_* $.}
		\setlength{\tabcolsep}{4mm}{
			\begin{tabular}{c|cc|cc|cc}
				\hline
				\hline
				$ j = m = k = 1 $ &
				\multicolumn{2}{c|}{$\mu=0.1$} &
				\multicolumn{2}{c|}{$\mu=0.2$} &
				\multicolumn{2}{c}{$\mu=0.3$} \\
				
				$ a_* $ & Re($\omega_*$) & -Im($\omega_*$)  & Re($\omega_*$) & -Im($\omega_*$) & Re($\omega_*$) & -Im($\omega_*$) \\
				\hline			
				0 & 3.149229 & 0.496878 & 3.153577 & 0.495903 & 3.160823 & 0.494282 \\
				0.1 & 3.161079 & 0.494751 & 3.165352 & 0.493797 & 3.172473 & 0.492209 \\
				0.5 & 3.019733 & 0.453793 & 3.023490 & 0.452984 & 3.029752 & 0.451636 \\
				0.8 & 2.793486 & 0.407248 & 2.796804 & 0.406553 & 2.802334 & 0.405395 \\
				1 & 2.629082 & 0.376645 & 2.632137 & 0.376012 & 2.637229 & 0.374956 \\
				2 & 1.951572 & 0.260459 & 1.953705 & 0.259971 & 1.957263 & 0.259156 \\
				3 & 1.130009 & 0.228908 & 1.130687 & 0.228718 & 1.132940 & 0.227006 \\
				4 & 0.991865 & 0.209493 & 0.992988 & 0.208406 & 0.995402 & 0.208082 \\
				5 & 0.906435 & 0.200252 & 0.907456 & 0.199819 & 0.909158 & 0.199097 \\
				6 & 0.849473 & 0.195660 & 0.850087 & 0.195541 & 0.851447 & 0.194943 \\
				7 & 0.809158 & 0.193197 & 0.809835 & 0.192888 & 0.810871 & 0.192622 \\
				\hline
				\hline
			\end{tabular}
			\label{table.1}
		}
	\end{table}

	To find possible instability, we also search for quasinormal modes for different values of angular quantum numbers $ m$  and $ j $, which are shown respectively in Fig.\eqref{fig2} and Fig.\eqref{fig3}. It is easy to see that there is no tendency to possible instability even for large $ m \text{ and } j $.  One can also check that all the quasinormal modes here  are not superradiant modes.	
	
	After carefully investigating the quasinormal modes at small values of $ \mu $, we find that no unstable modes exist. Then, we also make extensive searches for quasinormal modes for $ \mu > 1 $ cases. For smaller angular quantum numbers, it is found that the imaginary parts of the quasinormal modes tend to zero for large scalar masses. This long-living modes, called quasiresonances, are 
qualitatively the same as that found in four-dimensional Kerr black hole case \cite{Konoplya:2006br} and that in higher dimensional Schwarzschild black hole case  \cite{Zhidenko:2006rs}. For larger angular  quantum numbers, the imaginary parts of the quasinormal modes have a much slower tendency to zero. One such numerical result is shown in Fig.\eqref{fig4}. To calculate the cases with 
 large enough scalar masses $\mu$, one need a slow change of $\mu$ and long computation time. Also the method should be improved and the Nollert improvement can be used   \cite{Nollert:1993zz,Zhidenko:2006rs}.

	\section{Conclusion}\label{sec.6}
	In this work, we numerically study the stability of the six-dimensional singly rotating MP black hole against a massive scalar perturbation field.  
We compute the quasinormal modes of the massive scalar field using the continued fraction method. No matter slowly or rapidly rotating MP black holes, small or relatively large values of the scalar field masses, all found quasinormal modes are stable. Our result provides a direct numerical and complement analysis for that in Ref.\cite{Cardoso:2005vk} which gave a bound-state analysis for the stability of higher dimensional singly rotating MP black holes against massive scalar perturbations. 

It is also found that quasiresonances  with infinitely long lifetime appear in the fundamental quasinormal modes when the scalar masses $\mu$ become relatively large. In the four-dimensional Kerr black hole case, quasiresonances were also found for massive scalar fields\cite{Konoplya:2006br}. Furthermore, compared with the four-dimensional Kerr black hole case, scalar field modes in a six-dimensional singly rotating MP black hole have a faster damping. The similar property was also found for quasinormal modes of massive scalar fields in spherically symmetric Schwarzschild black holes \cite{Zhidenko:2006rs}.

	\section*{acknowledgements}
	We are grateful to R. A. Konoplya and A. Zhidenko for their useful comments. We also would like to thank Zhan-Feng Mai  and Lihang Zhou for useful discussions and suggestions. This work is partially supported by Guangdong Major Project of Basic and Applied Basic Research (No.2020B0301030008).

	
\end{document}